\documentclass[journal]{IEEEtran}

\usepackage{graphicx}
\usepackage[inkscapelatex=false]{svg}
\usepackage{verbatim}

\ifCLASSINFOpdf
\else
\fi
\usepackage{multirow, makecell}
\usepackage{booktabs}

\usepackage{siunitx}

\begin{document}
%
\title{The Spheres Dataset: Multitrack Orchestral Recordings for Music Source Separation and Information Retrieval}
%
%
%

\author{Jaime~García-Martínez,
        David~Diaz-Guerra,
        John Anderson,
        Ricardo~Falcón-Pérez,
        Pablo~Cabañas-Molero, Tuomas~Virtanen,~\IEEEmembership{Fellow,~IEEE,}
        Julio~J.~Carabias-Orti,
        and~Pedro~Vera-Candeas
\thanks{J. García-Martínez, P. Cabañas-Molero, J. J. Carabias-Orti, and P. Vera-Candeas are with the Universidad de Jaén, Spain (e-mail: \{jagarcim, pcabanas, carabias, pvera\}@ujaen.es).}
\thanks{John Anderson is with Odratek BV, Netherlands (e-mail: john@odratek.com).}
\thanks{D. Díaz-Guerra, R. Falcón-Pérez, and T. Virtanen are with Tampere University, Finland (e-mail: \{david.diaz-guerra, ricardo.falconperez, tuomas.virtanen\}@tuni.fi).}
\thanks{Manuscript received April 19, 2005; revised August 26, 2015.}}

%
%

\markboth{Journal of \LaTeX\ Class Files,~Vol.~14, No.~8, August~2015}%
{Shell \MakeLowercase{\textit{et al.}}: Bare Demo of IEEEtran.cls for IEEE Journals}

\maketitle

\begin{abstract}
This paper introduces The Spheres dataset, multitrack orchestral recordings designed to advance machine learning research in music source separation and related MIR tasks within the classical music domain. The dataset is composed of over one hour recordings of musical pieces performed by the Colibrì Ensemble at The Spheres recording studio, capturing two canonical works—Tchaikovsky’s Romeo and Juliet and Mozart’s Symphony No. 40—along with chromatic scales and solo excerpts for each instrument. The recording setup employed 23 microphones, including close spot, main, and ambient microphones, enabling the creation of realistic stereo mixes with controlled bleeding and providing isolated stems for supervised training of source separation models. In addition, room impulse responses were estimated for each instrument position, offering valuable acoustic characterization of the recording space. We present the dataset structure, acoustic analysis, and \textcolor{black}{different experiments} for orchestral family separation and microphone debleeding. Results highlight both the potential and the challenges of source separation in complex orchestral scenarios, underscoring the dataset’s value for benchmarking and for exploring new approaches to separation, localization, dereverberation, and immersive rendering of classical music.\end{abstract}

\begin{IEEEkeywords}
Music source separation, orchestral recordings, dataset, multitrack audio, microphone bleeding, debleeding, room impulse responses (RIRs), classical music, machine learning, immersive audio.
\end{IEEEkeywords}

\IEEEpeerreviewmaketitle

\vspace{-2mm}
\section{Introduction}
\IEEEPARstart{R}{ecent} advancements in music source separation (MSS) using artificial intelligence have been driven by the research community and the availability of benchmark challenges \cite{MUSDB18HQ,Mitsufuji2022music, Fabbro2024}. In professional music production, instruments and vocals are typically recorded on individual tracks, which are later combined during mixing. Although this process is often approximated as a simple sum of the isolated tracks, the final master undergoes complex, non-linear processing during mastering, making the resulting mix more than just an additive combination. Despite this simplification, assuming a linear mix remains effective in practice and does not significantly hinder model performance. MSS systems, therefore, aim to take a mixed audio signal as input and estimate the original constituent tracks, also known as stems; \textcolor{black}{for a comprehensive discussion on mixture model types and the specific complexities of the classical music domain, see~\cite{Burred2009}}.

In the last years, machine learning methods—particularly those driven by data—have become a central focus in music source separation research \cite{Cano2019}. Among these, deep neural networks have shown great promise, leading to notable improvements in separation accuracy. Supervised training of these models generally relies on datasets that contain isolated source tracks, which serve as ground truth references to guide learning. However, acquiring such data poses a major challenge due to copyright restrictions, and access to full multi-track recordings remains limited, as they are seldom released by artists. Despite these barriers, the research community has made considerable progress by curating and releasing multi-track datasets \cite{MUSDB18HQ, MTGMASSdb, Bittner2016MedleyDB2N, manilow2019cutting, ozer2023piano}, which have fueled advancements in separating sources across genres like pop and vocal music. Of particular relevance is the MUSDB18 dataset \cite{MUSDB18HQ,Mitsufuji2022music}, which serves as the foundation for standardized challenges where algorithms are benchmarked under consistent conditions to separate stems such as vocals, drums, bass, and accompaniment. These efforts have led to the emergence of cutting-edge models that have significantly pushed the boundaries of what is possible in source separation \cite{Fabbro2024,defossez2019music, spleeter2020, defossez2021hybrid, rouard2022hybrid, luo2023music,Tong2024,Lu2024,chen_dttnet_2024}.

However, music source separation in the orchestral domain has received significantly less attention compared to popular music and the several approaches that have been presented \cite{Miron2017Monaural, Slizovskaia2021, ozer2024source} suffer  challenges such as: (i) Limited availability of datasets suitable for training deep learning models.
(ii) Classical ensembles typically feature a greater number of instruments.
(iii) Certain instruments within an ensemble possess similar timbral characteristics, making them more difficult to differentiate (e.g., violin and viola).
(iv) Recordings are commonly captured with all musicians performing together in the same space, introducing room acoustics into the audio signals.

Training supervised source separation models requires clean, isolated source tracks to serve as ground truth during learning. However, acquiring such recordings for orchestral or ensemble music presents substantial challenges. Unlike popular music, where individual performers can be recorded separately using a metronome or backing track, classical ensembles are typically recorded as a group in a single take to preserve natural synchronization and enable joint musical expression. This practice, rooted in the inherently collective nature of rehearsals and performances, leads to recordings that contain significant bleed from other instruments even when each instrument is close miked. Moreover, the large number of instruments involved in orchestral settings makes isolated, instrument-by-instrument recording both logistically complex and musically unnatural. In fact, several limited databases of real-world recordings have been developed for classical ensembles (see \cite{Garcia2025} for a detailed overview). Small-ensemble efforts include the TRIOS dataset \cite{TriosDataset}, which provides bleed-free separated tracks for chamber music trios, and the Bach10 dataset \cite{BachChorals}, which offers bleed-free signals of four-part chorales recorded with musicians in isolation. The URMP dataset \cite{URMPdataset} features 44 pieces for small ensembles (2 to 5 instruments) created from individually recorded but temporally aligned performances. \textcolor{black}{Similarly, the ChoraleBricks dataset \cite{Balke_ChoraleBricks_TISMIR} provides a modular framework of ten four-part chorales, where various wind instruments were recorded in isolation to allow for the flexible synthesis of ensembles with different instrumentations}. For larger orchestral material, the Aalto Anechoic Orchestra dataset \cite{Anechoicdataset} provides approximately 10 minutes of anechoic recordings of individual instruments. More recently, the Operation Beethoven project \cite{OperationBeethovenB} released around 10 minutes of bleed-free isolated recordings of orchestral sections, captured in a concert hall to preserve natural reverberation.

Unfortunately, the scarcity of suitable training material has significantly constrained progress in supervised
source separation research for classical music. To mitigate this, current systems rely on data augmentation or are trained on synthesized audio and subsequently evaluated on real recordings. 

For data augmentation, several approaches have been proposed for classical-music \textcolor{black}{MSS} \cite{chiu2020mixing, kim2023study, ozer2024source}, but none targets separation of all sections in a full orchestra. Alternatively, \cite{miron2017generating} assumes that the score is available and uses multiple synthetic renditions of the piece to train the MSS system. Recently, \cite{tunturi2025} presented a score-informed MSS model that uses only score information to generate separation masks, demonstrating its ability to generalize from synthetic to real data. The work in \cite{Manilow2020HierarchicalMI} introduces a novel approach that leverages the hierarchical relationships between musical instruments to achieve more flexible and context-aware MSS, improving performance with limited training data.

Regarding the synthetic, Sarkar et al. \cite{sarkar2022ensembleset} developed a bleed-free synthesized dataset called EnsembleSet and used it to train a duet source separation model based on a dual-path transformer architecture. For evaluation, they extracted duet segments from real recordings in the URMP dataset \cite{URMPdataset}. A recent study introduced SynthSOD \cite{Garcia2025}, a synthesized dataset that offers a more balanced representation of instruments compared to Ensembleset. To evaluate the effectiveness of this new dataset, García-Martínez et al. trained four separate X-UMX models \cite{Sawata2021}, each dedicated to one of the following instrument families: (i) strings, (ii) woodwinds, (iii) brass, and (iv) percussion. Evaluation on the URMP dataset resulted in generally low Source-to-Distortion Ratio (SDR) scores, indicating limited separation performance. These outcomes highlight that MSS involving multiple sources and based on real, recorded ensembles continues to be a difficult and largely unsolved problem in the context of classical music. Recent initiatives have begun to address some of these challenges. For example, within the Cadenza project\footnote{https://cadenzachallenge.org/}, a dedicated task has been developed to rebalance classical music to enhance music perception for individuals with hearing loss. Although the repertoire is limited to small ensembles of woodwind instruments \cite{Roa2024}, this represents, to the best of our knowledge, the first challenge specifically focused on classical music in this context, further highlighting the growing attention this domain is beginning to receive.

In this paper, we present The Spheres dataset, the first publicly available\footnote{https://doi.org/10.5281/zenodo.17347681} multitrack orchestral resource capturing classical music through a comprehensive set of microphone signals, including ambient, main and close spot microphones. This dataset was developed within the framework of the REPERTORIUM project\footnote{https://repertorium.eu/} and is specifically designed to support the development of machine learning methods for source separation and related MIR tasks in the classical music domain. The dataset comprises over one hour of music recordings performed by the Colibrì Ensemble\footnote{https://www.colibriensemble.it/}
(Chamber Orchestra of Pescara, Italy) recorded at The Spheres recording studio\footnote{https://www.the-spheres.com/}, a venue selected for its optimal acoustical conditions and professional infrastructure. Unlike previous datasets, in The Spheres each instrument/part was recorded in isolation (one at a time) while all microphones captured the performance, yielding leakage-free stems per instrument and microphone position. Realistic ensemble mixes with controlled bleed are then obtained by summing the isolated contributions across parts and microphones. This design captures realistic orchestral performances with individual instrument spot microphones while preserving the natural acoustics and ensemble coordination inherent to classical music. By combining full ensemble mixtures derived from isolated stems with spatial calibration signals, The Spheres dataset opens new avenues for research in supervised separation, source localization, dereverberation, and immersive audio rendering for classical music.

In addition, \textcolor{black}{we present different MSS experiments.} First, we evaluate monaural family separation (strings, woodwinds, brass, percussion) from the \textcolor{black}{stereo mixture}. Second, we address a production-oriented task—debleeding of section spot microphones—by training lightweight single-branch models that enhance a target section from its own spot channel while suppressing orchestral leakage. These \textcolor{black}{experiments} expose both the promise and the difficulty of orchestral MSS: measurable interference reduction is achievable, yet generalization across repertoire and setups remains challenging. We release all materials and scripts so that future works can compare methods under consistent conditions, explore multichannel and score-informed models, and exploit the included RIRs and solo material to bridge synthetic–to–real gaps when targeting complex orchestral signals.

\section{Recording the dataset}

This section describes the process of creating The Spheres dataset, including the musical repertoire performed, the recording sessions, and the microphone setup employed. These details provide the basis for understanding the structure and properties of the released material.

\subsection{Music material}

The recordings include two pieces of music, as well as solo material of each instrument to allow potential training data generation.
The first piece of music, Romeo and Juliet  (overture-fantasia), TH 42, is an overture by Pyotr Ilyich Tchaikovsky, whose final version was completed in 1880. The duration of the overture is 22~min~38~s. The overture is scored for an orchestra comprising piccolo, 2 flutes, 2 oboes, English horn, 2 clarinets (in A), 2 bassoons, 4 horns (in F), 2 trumpets (in E), 3 trombones, tuba, 3 timpani, cymbals, bass drum, harp, violins I, violins II, violas, cellos, and double basses.

The second piece of music is Symphony No. 40 in G minor, K. 550 (revised version with clarinets) written by Wolfgang Amadeus Mozart in 1788. The total duration is 30~min~16~s. The symphony is scored for flute, 2 oboes, 2 clarinets, 2 bassoons, 2 horns, and strings. The work is in four movements, in the usual arrangement for a classical-style symphony (fast movement, slow movement, minuet, fast movement) --- Molto Allegro, Andante, Minuetto. Allegretto – Trio, Finale.

The above pieces of music were selected because of their wide appreciation in the Western culture, as well as because the copyright of the compositions has expired, allowing their royalty-free use.

In addition, each section of instruments played a full chromatic scale across the instrument’s natural range with different dynamics (quiet/loud) and playing techniques (legato, pizzicato, staccato, etc). An individual instance of each instrument played by the section leader also played solo scales with different dynamics, as well as free choice solo content.

\subsection{Recording setup}
\label{sec:recoding_setup}

\begin{figure}
    \centering
    \includegraphics[width=\linewidth]{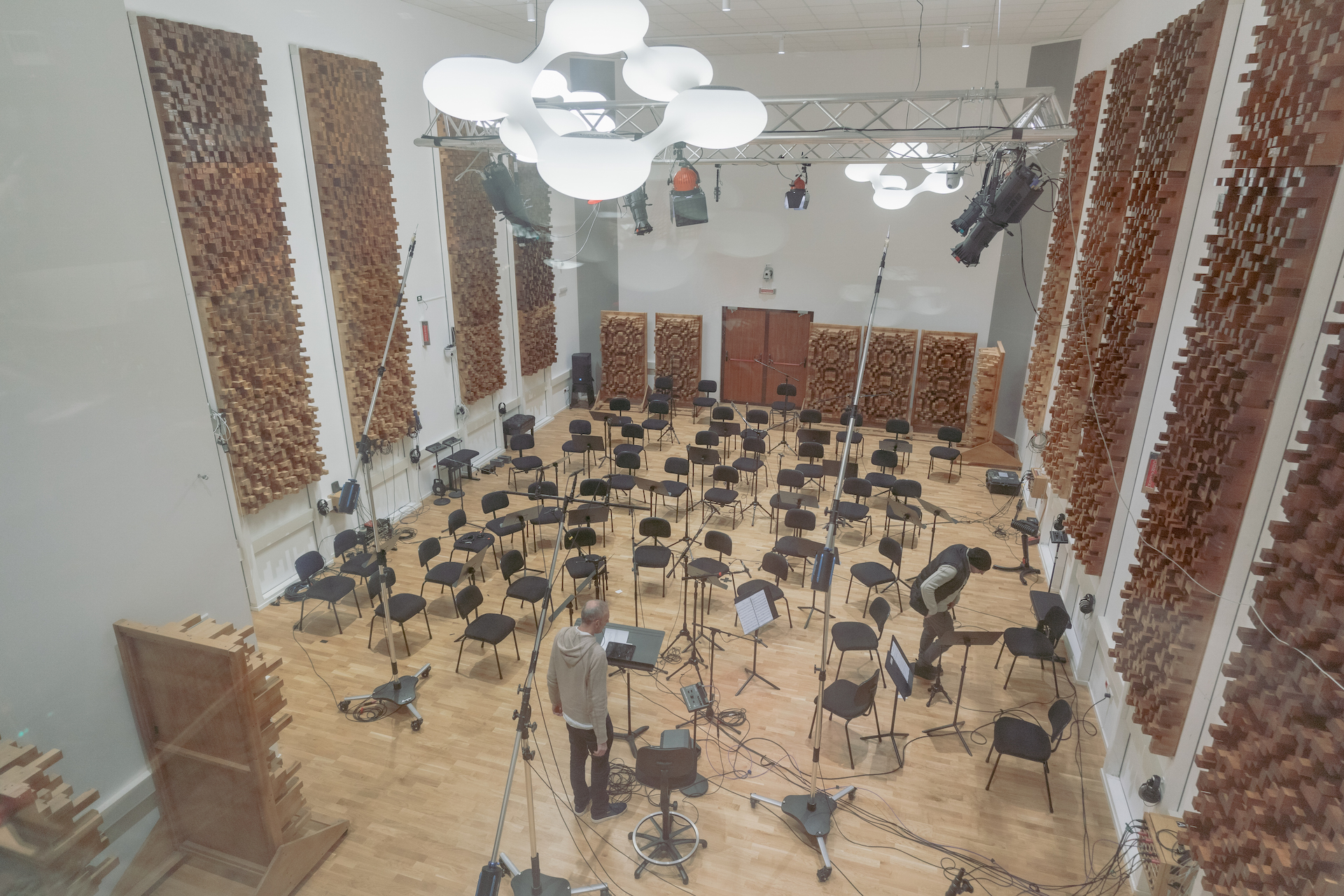}
    \caption{A photo of the studio used for the recordings. Each musician was allocated a seat in the studio, which placement was following the typical placements when playing orchestral music.}
    \label{fig:recordingroom}
\end{figure}

The music was recorded in the Main Room of The Spheres Recording Studio, whose size is 15 x 9 x 8 meters. The high-quality acoustic design of the room includes diffusors, wedge bass traps, sound-absorbent dropped ceiling, and floating walls. See Section \ref{sec:analysis} for acoustic measurements of the room. Figure \ref{fig:recordingroom} presents an overview of the recording room and the recording setup.

\begin{figure}
    \centering
    \includegraphics[width=\linewidth]{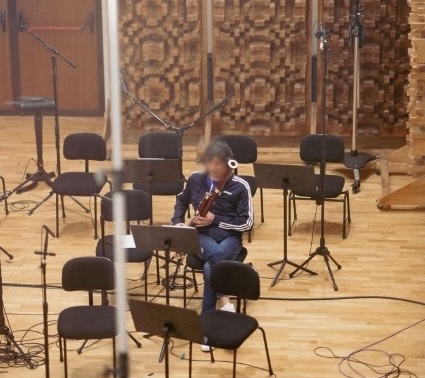}
    \caption{A photograph illustrating a session of recording one bassoon line.}
    \label{fig:bassoon-example}
\end{figure}

\begin{table}
    \centering
    \caption{The number of separately recorded lines per each instrument class for the two pieces of recorded music.}
    \begin{tabular}{|l|c|c|}
    \hline
\textbf{Instrument class}    & Symphony No. 40 & Romeo and Juliet \\ \hline
Bass & 1 & 1 \\ \hline
Bass drum & & 1	 \\ \hline
Bassoon &2 & 2	 \\ \hline
Cello &1 & 2	 \\ \hline
Clarinet &2 & 2	 \\ \hline
Cymbals & & 1	 \\ \hline
English horn & & 1	 \\ \hline
Flute &1 & 2	 \\ \hline
Harp	& & 1 \\ \hline
Horn &2 & 4	 \\ \hline
Oboe &2 & 2	 \\ \hline
Piccolo & & 1	 \\ \hline
Timpani	& & 1 \\ \hline
Trombone & & 3	 \\ \hline
Trumpet & & 2	 \\ \hline
Tuba & & 1	 \\ \hline
Viola &1  & 2	 \\ \hline
Violin 1 &1 & 2	 \\ \hline
Violin 2 & 1 & 2    \\ \hline
    \end{tabular}
    \label{tab:instrumentlines}
\end{table}

\begin{figure*}
    \centering
    \includegraphics[width=0.6\linewidth]{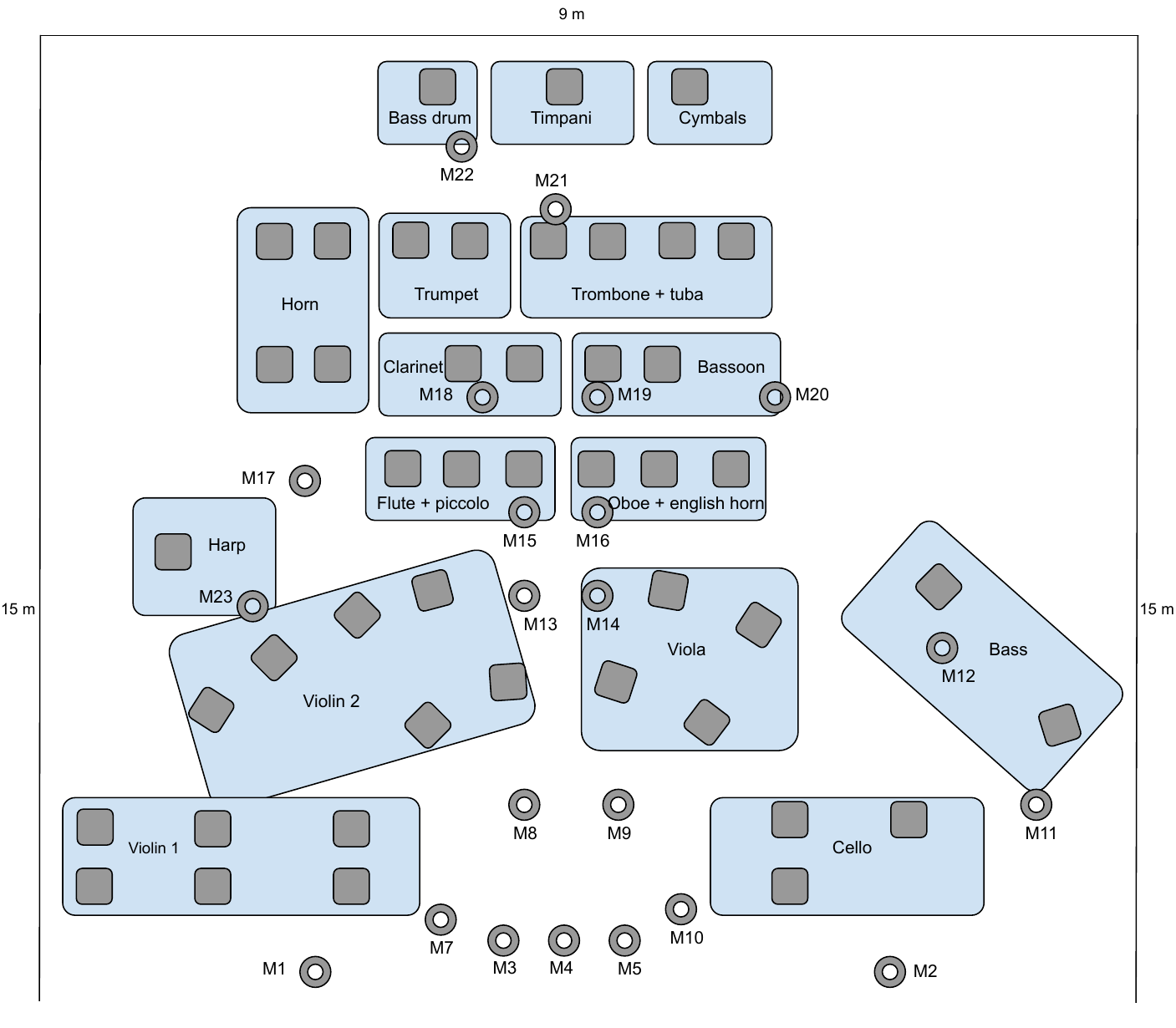}
    \caption{The approximate placement of instruments  and microphones (indicated by M\#) in the recording room. Each rounded square indicates the seat of a musician.}
    \label{fig:instrument-positions}
\end{figure*}

Each part of each instrument was played in isolation, to enable capturing instrument-specific tracks. To achieve this, a separate recording session was organized for each instrument at a time (i.e., separate recording session for each part of violin 1, violin 2, viola, each line of horns, etc.). The number of separately recorded lines per each instrument class for the two pieces of recorded music is given in Table \ref{tab:instrumentlines}.
The musicians wore headphones and listened to reference tracks, so that the separately recorded parts were synchronous in time. As the reference tracks, the recordings of New York Philharmonic \cite{tchaikovskyyoutube} and Boston Symphony Orchestra \cite{mozartyoutube} conducted by Leonard Bernstein were used. A conductor was conducting musicians at each recording session, to create a more natural scenario. Figure \ref{fig:bassoon-example} illustrates an example of the bassoon recording session. %
The locations of musicians in the recording room followed their typical placement in orchestras, as visualised in Figure \ref{fig:instrument-positions}. The seats of musicians were kept fixed between the recording sessions. 

The recordings were captured using 23 microphones, which included two ambient microphones, three main microphones, and 18 close microphones for individual instruments. All the microphones were used to capture all the instruments, to allow capturing realistic acoustic propagation from each instrument to all the microphones. The list of microphones is given in Table \ref{tab:channel_instrument_mic}. The placement of microphones is illustrated in Figure \ref{fig:instrument-positions}. Microphone \#6 is a close spot microphone that was moved to be close to each instrument when recording the part of that instrument; the goal of this microphone was to capture a drier sound than the instrument/section microphones \textcolor{black}{and could be used, for example, as source signals in acoustic simulations of different rooms}. The recordings were done in two stages, with some differences in the microphones used in them. The instruments recorded in Session 1 included 6 violins, 4 violas, 3 cellos, and 2 basses, and Session 2 included the rest of the instruments or sections. The piccolo player shared the seat with one of the flute players, and the English horn player shared the seat with one of the oboe players. 

All the microphones were connected to DAD AX32 AD-converter to allow synchronous capture. The recording setup, including the seats and microphones, were reinstalled between the two stages. The locations were kept similar between the sessions, but because of the installation procedure there were small deviations in the locations.

\begin{table}[ht]
    \centering
    \caption{The channel ID of each instrument and microphones used to record them.}
    \begin{tabular}{|c|l|l|}
    \hline
    \textbf{ID} & \textbf{Location}           & \textbf{Microphone (session 1 / 2)} \\ \hline
    M1                   & Amb L                         & Earthworks QTC1 / Schoeps MK2        \\ \hline
    M2                   & Amb R                         & Earthworks QTC1 / Schoeps MK2        \\ \hline
    M3                   & Mains L                       & DPA 4006         \\ \hline
    M4                   & Mains R                       & DPA 4006         \\ \hline
    M5                   & Mains C                       & DPA 4006         \\ \hline
    M6                   & Close spot            & Schoeps MK4          \\ \hline
    M7                   & Violin 1 section              & Schoeps MK4          \\ \hline
    M8                   & Violin 2 section              & Schoeps MK4          \\ \hline
    M9                   & Viola section                 & Schoeps MK4          \\ \hline
    M10                  & Cello section                 & Schoeps MK4          \\ \hline
    M11                  & Double bass 1                 & Schoeps MK4          \\ \hline
    M12                  & Double bass 2                 & Schoeps MK4          \\ \hline
    M13                  & Flute / piccolo section         & Schoeps MK4 / DPA 4011        \\ \hline
    M14                  & Oboes     & Schoeps MK4 / DPA 4011        \\ \hline
    M15                  & Clarinets                     & Schoeps MK4 / DPA 4011      \\ \hline
    M16                  & Bassoons section              & Schoeps MK4 / DPA 4011         \\ \hline
    M17                  & Horn section                  & Schoeps MK4 / Schoeps V4         \\ \hline
    M18                  & Trumpets section              & Schoeps MK4 / Schoeps V4          \\ \hline
    M19                  & Trombones section              & Schoeps MK4 / Schoeps V4          \\ \hline
    M20                  & Tuba              & Schoeps MK4 / Schoeps V4          \\ \hline
    M21                  & Timpani              & Schoeps MK4 / 4015          \\ \hline
    M22                  & Bass drum  and cymbals              & Schoeps MK4 / Schoeps MK21          \\ \hline
    M23                  & Harp              & Schoeps MK4           \\ \hline    
    \end{tabular}
    \label{tab:channel_instrument_mic}
\end{table}

\textcolor{black}{Apart from the raw recordings from each microphone described above, the dataset includes a Stereo Mix created by a professional audio engineer. This mix was produced by combining signals from the main and spot microphones (Mains L and R channels) to create a realistic balance that simulates typical classical music production.
The main mixture refers to the linear sum of the individually recorded stems, preserving natural instrument bleed between spot and main microphones. In contrast, the Stereo Mix includes additional post-processing, such as artificial reverberation and limiting, resulting in a polished, production-ready stereo recording suitable for listening or evaluation.}

Several stems, defined as the contribution of each individually recorded instrument and its corresponding parts to the main mixture, were also produced to allow various kinds of source separation experiments based on the main mixture. The sum of all the stem signals equals the main mixture. The stem signals can be summed in various ways to be used in source separation experiments. For example, summing all the lines of each instrument class will create instrument-specific reference signals, and summing all the lines of each section (strings, woodwinds, brass, and percussion), will create section-specific reference signals.

Dedicated measurements to capture the acoustic characteristics of the recording space were performed. Specifically, exponential sine sweeps (ESSs) \cite{farina2007advancements} were fed to the recording environment using a Neumann KH120 II studio monitor placed at the position of the first chair of each section, with its height adjusted to match the typical playing height of the corresponding instrument and oriented facing the main microphone array. In addition, hand claps were recorded from the same positions as the ESS. The resulting signals were recorded by all microphones in the setup and used to estimate room impulse responses (RIRs) for multiple source–receiver pairs, providing a detailed characterization of the hall’s spatial acoustics.

To estimate the RIRs from the recorded ESS signals, an inverse filter was derived from the reference sweep signal so that its convolution with it approximates a delayed Dirac delta. Convolving each recorded microphone signal with this inverse filter recovers the RIR, while non-linear artifacts appear at earlier times and can be removed through time-windowing. Due to corruption of the reference sweep signal, a modified inverse filter was computed in the frequency domain; full details of this procedure are provided in Appendix~\ref{appendix:ESS}.

\textcolor{black}{After the recording sessions, several failures were observed in the recordings, resulting in missing the double bass signals in the main right microphone, the timpani, bass drum, and cymbals signals in the clarinet microphone, and the flute signal in the close-spot microphone. The published dataset contains files for those signals, but they are completely silent.}

\section{File structure of the dataset}

In order to facilitate code reusability and cross-dataset comparisons, we have organized The Spheres dataset following the same structure as EnsembleSet \cite{sarkar2022ensembleset} and SynthSOD \cite{Garcia2025}. Each music piece is stored in a dedicated folder containing subfolders for every microphone (including the stereo mix), with each subfolder holding an audio file for each signal of each single instrument. One important difference from the previously mentioned datasets is that The Spheres may contain more than one file for some instruments, corresponding to separately recorded lines (see Table \ref{tab:instrumentlines}). Another difference is that the audio files in The Spheres are sampled at \SI{48}{kHz}, instead of \SI{44.1}{kHz}. 

By summing all audio files within a folder, the signal corresponding to that microphone (or the stereo mix) can be reconstructed as if all instruments had been performed simultaneously. This approach is not applicable to the close spot microphone which, as described in Section~\ref{sec:recoding_setup}, was repositioned during the recordings; thus, summing its signals is not meaningful. The dataset also contains scales and solos for each instrument as a third type of music piece, although combining these signals results in a noisy, non-musical outcome and they have differing durations.

We provide some metadata files similar to the ones included in the SynthSOD dataset \cite{Garcia2025}, containing basic information about the instruments present in each piece. In addition, we also publish an independent version including only the stereo mix, so that users interested solely in these signals do not need to download the full multichannel version of the dataset.

To complement the multi-channel recordings, we provide estimated RIRs for each instrument position. The estimated RIRs are stored as NumPy binary files (``.npy''), each containing a two-dimensional array of shape \([M, N]\), where \(M\) is the number of microphones in the recording setup and \(N\) is the number of samples. The filename encodes the source position of the excitation signal. For example, \texttt{source\_Vln\_1.npy} contains the RIRs measured when the source was located at the Violin~1 position. The microphone channel order is consistent across all \texttt{.npy} files.  

\begin{figure}
    \centering
    \includegraphics[width=\linewidth]{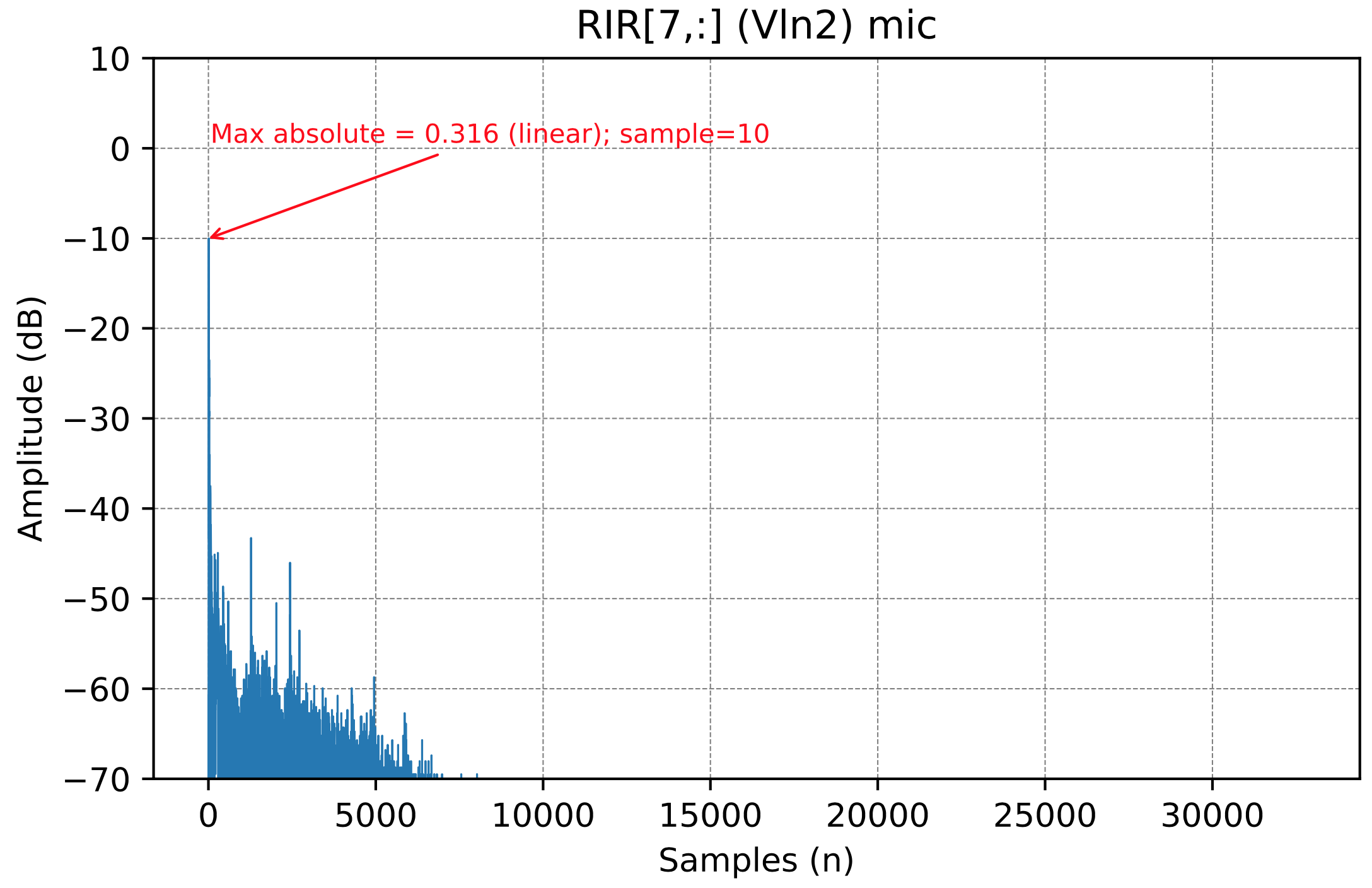}
    \caption{Estimated RIR for the Violin 2 microphone with the source located at the Violin 2 position. The plot shows the RIR in dB as a function of sample index, including annotations for the microphone index (7) and name (Vln2).}
    \label{fig:rir_plots}
\end{figure}

To facilitate exploration of the dataset, each \texttt{.npy} file is accompanied by a \texttt{.pdf} document containing plots visualizing the corresponding RIRs across all microphones (see Figure \ref{fig:rir_plots}). These plots display the RIRs amplitude in \textit{decibels (dB)} as a function of \textit{sample index}. Each plot is annotated with the microphone index and name.

In addition, we provide the original recordings from all microphones of the claps and sweeps at each instrument position, enabling researchers to perform their own RIR estimations if desired.

\vspace{-3mm}
\section{Analysis and properties of the dataset}
\label{sec:analysis}

In this section, we provide a quantitative analysis of The Spheres dataset in order to characterize its musical content and recording conditions. We first examine the distribution of instrument activity and polyphony levels across the two orchestral pieces, highlighting differences in instrumentation and texture. Next, we evaluate the signal quality of the recordings by reporting signal-to-distortion ratios (SDR) and quantifying the amount of bleeding between microphones, with a particular focus on section microphones. Finally, we analyze the acoustics of the recording space using the captured RIRs, from which reverberation time and clarity metrics are derived. Together, these analyses provide a comprehensive view of the dataset’s properties and the challenges it poses for music source separation and related tasks.

\vspace{-2mm}
\subsection{Instrument times and polyphony}

\begin{figure}
    \centering
    \includegraphics[width=0.85\linewidth]{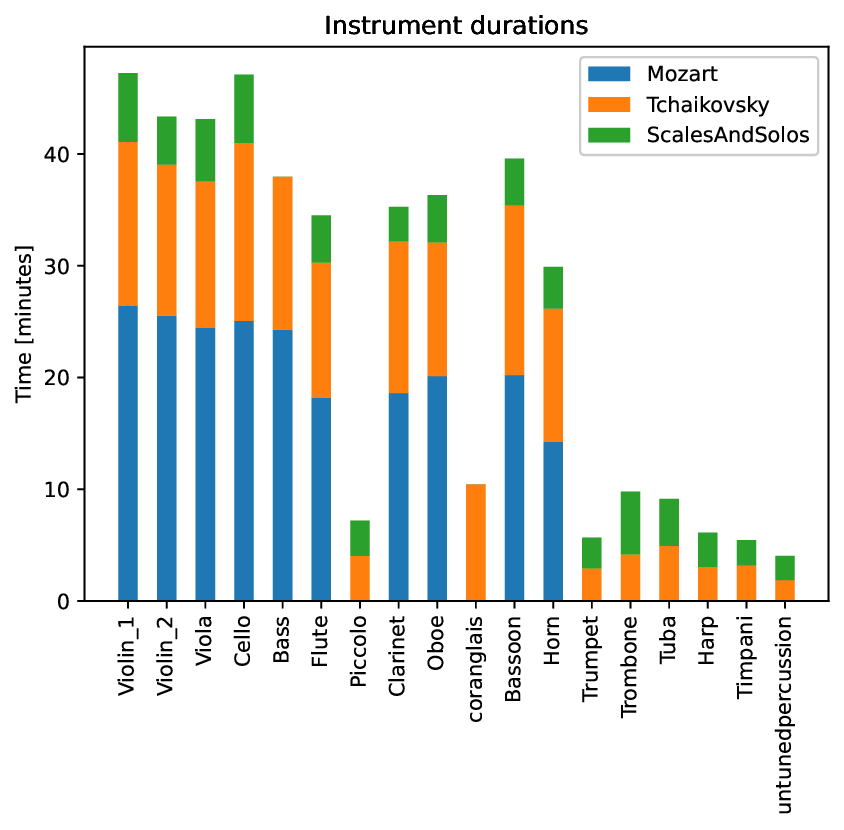}
    \caption{Time played by every instrument in The Spheres dataset.}
    \label{fig:intrument_times}
\end{figure}

\begin{figure}
    \centering
    \includegraphics[width=0.85\linewidth]{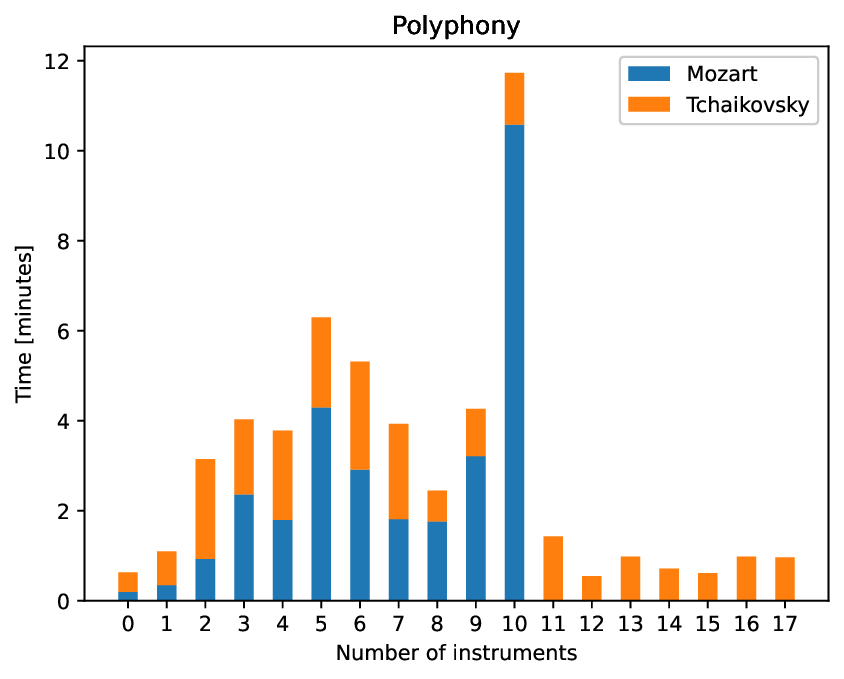}
    \caption{Time of every polyphony level in The Spheres dataset.}
    \label{fig:polyphony}
\end{figure}

Figure \ref{fig:intrument_times} shows the time played by each instrument in the dataset. \textcolor{black}{While the Mozart piece provides over 20 minutes of audio for all string and woodwind sections, it lacks piccolo, English horn, percussion, and brass instruments (with the exception of the horn). In contrast, the Tchaikovsky piece has a shorter duration but offers a wider variety of instrumentation.}

In Figure \ref{fig:polyphony}, we can see the time of each polyphony level in the dataset. During around one third of its duration, the Mozart piece has all its instruments playing at the same time, whereas the polyphony levels in the Tchaikovsky piece are more evenly distributed, with the different instruments appearing and disappearing during the piece, and there is not a single moment when all the instruments are playing at the same time.

In order to compute both statics, the signals of every instrument at the main left microphone were split into 1-second non-overlapping frames, and instruments were considered to be active in those frames whose energy were above a \SI{-30}{dB} threshold with respect to its more energetic frame. This is the same criteria that was used to exclude de silent frames from the evaluation of the experiments presented in section \ref{sec:experiments}.

\vspace{-2mm}
\subsection{Signal-to-distortion ratio and bleeding in the mixtures}

\begin{table*}
    \sisetup{table-format=2.1, table-auto-round}
    \centering
    \caption{Signal-to-distortion ratio (SDR) of every instrument in the Stereo Mix, the main and amb microphones, and in their section close microphone.}
    \begin{tabular}{llSSSSSSS}
    \toprule
    \multirow{2}{*}{Piece}          & \multirow{2}{*}{Source}   & \multicolumn{7}{c}{SDR [dB] at microphone}                                                \\
                                                                \cmidrule(lr){3-9} 
                                    &                           & {Stereo Mix}  & {Main L}  & {Main C}  & {Main R}  & {Amb L}   & {Amb R}   & {Section}     \\
    \midrule
    \multirow{10}{*}{Mozart}        & Violin I                  & -12.50        & -10.64    & -11.21    & -10.27    &  -7.60    &  -8.87    & -2.44         \\
                                    & Violin II                 & -14.90        & -11.31    & -11.55    & -11.51    & -15.25    & -16.35    & -7.01         \\
                                    & Viola                     & -10.50        & -12.26    & -11.41    & -11.53    &  -9.77    &  -8.99    & -2.81         \\
                                    & Cello                     &  -8.33        & -10.35    & -10.52    &  -8.79    &  -8.07    &  -7.13    & -2.45         \\
                                    & Bass                      &  -2.12        &  -9.02    &  -9.99    &           &  -7.40    &  -5.90    &  5.70         \\
                                    & Flute and piccolo         & -21.66        & -14.37    & -14.35    & -13.84    & -15.56    & -15.89    & -1.75         \\
                                    & Clarinet                  &  -9.78        &  -6.73    &  -6.03    &  -5.73    & -10.25    & -10.86    &  2.76         \\
                                    & Oboe and cor anglais      & -19.73        & -11.82    & -11.58    & -10.87    & -14.32    & -14.20    & -3.76         \\
                                    & Bassoon                   & -12.70        & -10.34    & -10.92    &  -9.78    & -12.59    & -14.34    & -0.19         \\
                                    & Horn                      &  -9.01        &  -3.53    &  -2.96    &  -2.60    &  -5.25    &  -6.23    & 11.86         \\
    \midrule
    \multirow{16}{*}{Tchaikovsky}   & Violin I                  & -13.59        & -11.88    & -12.82    & -12.57    &  -9.03    & -10.12    & -4.64         \\
                                    & Violin II                 & -12.63        & -12.05    & -13.12    & -13.39    & -16.16    & -16.57    & -8.21         \\
                                    & Viola                     & -13.41        & -13.67    & -13.50    & -13.70    & -10.87    & -10.25    & -5.61         \\
                                    & Cello                     &  -8.11        & -10.06    &  -9.77    &  -8.80    &  -6.38    &  -5.76    & -1.99         \\
                                    & Bass                      &  -8.09        & -15.47    & -16.51    & -10.66    & -13.33    & -12.37    & -0.68         \\
                                    & Flute and piccolo         & -18.43        & -11.49    & -12.13    & -11.96    & -13.95    & -13.93    & -2.54         \\
                                    & Clarinet                  &  -9.73        & -10.07    &  -3.82    &  -9.41    & -12.80    & -11.56    & -1.25         \\
                                    & Oboe and cor anglais      & -14.68        & -10.71    & -10.95    & -10.97    & -13.34    & -12.52    & -4.54         \\
                                    & Bassoon                   &  -9.82        & -11.39    & -12.13    & -11.36    & -13.49    & -14.95    & -0.93         \\
                                    & Horn                      &  -5.13        &  -4.06    &  -3.82    &  -3.52    &  -5.30    &  -4.91    &  9.64         \\
                                    & Trumpet                   & -13.35        &  -8.72    &  -8.00    &  -8.46    & -10.17    & -10.82    & -1.84         \\
                                    & Trombone                  & -13.05        &  -9.91    &  -9.02    & -10.87    & -11.25    & -12.25    & -5.90         \\
                                    & Tuba                      & -13.18        & -11.02    & -11.29    &  -8.80    & -11.37    & -14.10    & -3.96         \\
                                    & Harp                      & -12.50        &  -7.65    &  -7.68    & -13.35    &  -9.10    & -11.48    & -0.84         \\
                                    & Timpani                   &  -7.02        & -11.75    & -11.95    &  -8.93    &  -8.10    &  -8.92    & -1.43         \\
                                    & Untuned percussion        &  -5.43        &  -9.25    &  -8.77    &  -8.93    &  -6.37    &  -5.81    & -4.28         \\
    \bottomrule
    \end{tabular}
    \label{tab:original_sdr}
\end{table*}

\begin{figure}
    \centering
    \includegraphics[width=0.9\linewidth]{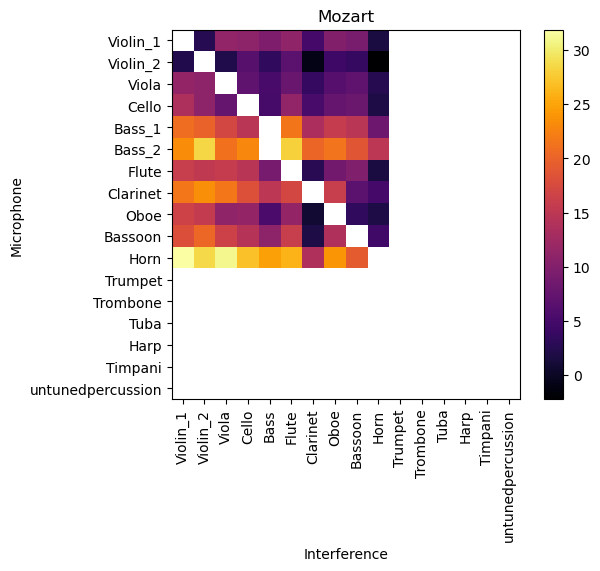}
    \includegraphics[width=0.9\linewidth]{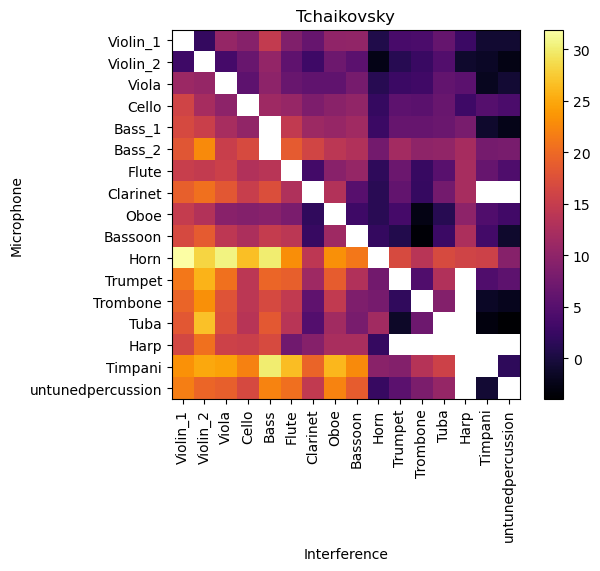}
    \caption{Signal-to-interference ratio [dB] for the main instrument/section of every section microphone against the each interference source.}
    \label{fig:bleeding_matrices}
\end{figure}

In Table \ref{tab:original_sdr} we can see the SDR of every instrument/section in the stereo mix, the main and amb microphones, and in their section close mic. \textcolor{black}{It should be noted that, since there are no artifacts or distortions in the original mix, the SDR here is equivalent to the signal-to-interferences ratio (SIR).} As done with the evaluation of the experiments presented in section \ref{sec:experiments}, the SDR of every instrument was computed first in 1-second non-overlapping frames and then aggregated by computing the median of those frames in which the instrument was active. \textcolor{black}{The value of the SDR of the Bass in the Main R microphone is missing since its signal was not captured due to a failure during the recordings (see section \ref{sec:recoding_setup}).}

From these results, it is clear how challenging the task of instrument/section separation from the stereo mixture or the main microphones of a symphonic orchestra is, with most sources having SDRs around \SI{-10}{dB}. As we could expect, the best SDRs for every instrument/section are obtained in their dedicated microphone; however, those still contain a huge amount of bleeding from the other instruments of the orchestra and most of them still have a negative SDR.

In order to have a better knowledge about the bleeding in the section microphones, we have computed, for each of them, the ratio between the energy of their target instrument/section and every interference source, and represented it as a matrix for every music piece of the dataset in Figure \ref{fig:bleeding_matrices}. \textcolor{black}{Results indicate that section microphones with superior SDR generally exhibit a better ratio against all interferences; however, some notable one-to-one interactions are also observable.}

For example, we can see how the viola section creates a higher bleeding in the violin II microphone than in violin I, which makes sense considering how they were placed in the room (see Figure \ref{fig:instrument-positions}). White cells correspond to cases where computing this ratio did not make sense (i.e., when the interference and the target of the microphone is the same or they do not overlap in time) or cases that could not be computed because of failures during the recording (i.e., the interference of the timpani and the untuned percussion on the clarinet microphone, whose signals were not recorded).

\subsection{Acoustic Analysis}

To understand the acoustics of the room we analyzed the RIRs that were captured as mentioned in section \ref{sec:recoding_setup}. While the recording procedure does not fully follow the standard (in particular, the use of non omnidirectional loudspeakers), it provides some useful information about the acoustics of the space. Here we focus on two acoustic parameters computed from the impulse responses. First, we show the clarity index C50, which is defined as the ratio of energy arriving in the first 50 milliseconds of the impulse to the energy arriving later. This reflects the balance between early sound and late reverberation, and it is highly correlated with closeness to the source, speech intelligibility and perceived ease of listening. Next, we compute the reverberation time using T30, which represents the time it would take for sound energy to decay by 60 dB. To avoid the influence of the noise floor, this is estimated by extrapolating the decay between the $-5$ and $-35$ dB levels. All acoustic parameters are computed according to the standard \cite{ISO3382-1:2009}. For reverberation time metrics, we notice some two noticeable patterns. First, there are some double slope decays, in mid to high frequencies, in most positions. This is subtle but present, and most likely due to the very high ceilings the uneven distribution of acoustic treatment (mostly concentrated at ground level). Secondly, there is a small but noticeable variance of acoustic parameters across the space. 

\begin{figure}
    \centering
\includegraphics[width=0.89\linewidth]{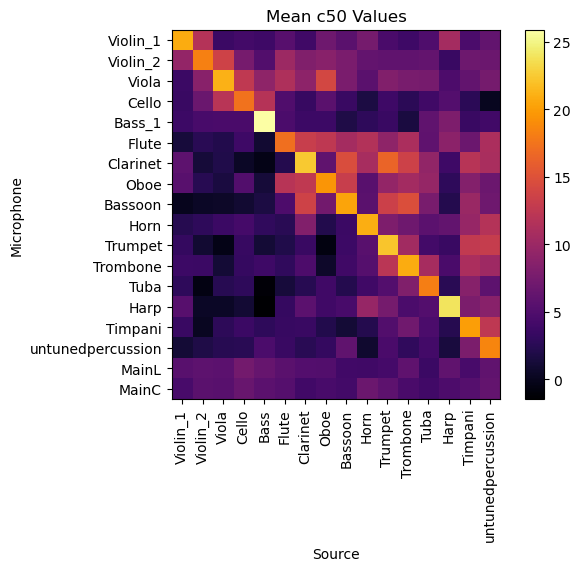}
    \caption{Distribution of C50}
    \label{fig:c50}
\end{figure}

\begin{figure*}
    \centering
    \includegraphics[trim={0cm 0cm 1cm 1.1cm}, clip, width=0.8\linewidth]{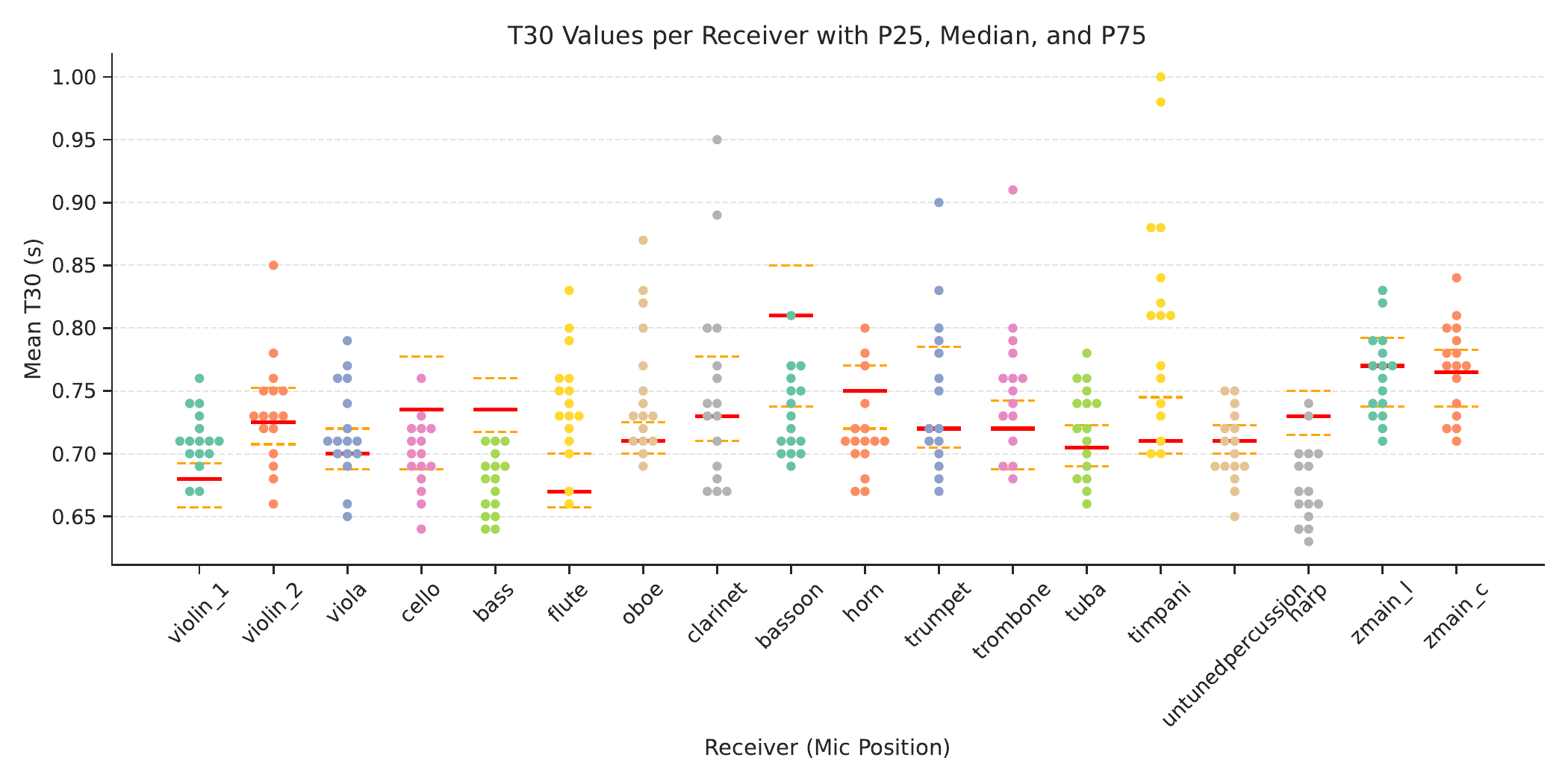}
    \caption{Swarmplot of T30 values for each receiver (microphone position). Solid and dashed lines indicate the median and the 25th and 75th percentiles, respectively.}
    \label{fig:t30}
\end{figure*}

Figure \ref{fig:c50} shows the C50 of all the source/receiver pairs. As could be expected, the matrix presents a strong diagonal, since the section microphones are closer to their corresponding section and steered towards them. Out of the diagonal, we can also observe several high values that correlate with the lower SIR values in Figure \ref{fig:bleeding_matrices} since, as can be seen in Figure \ref{fig:instrument-positions}, usually correspond to interfering instruments that are behind the target instrument of the microphone, and therefore have a strong direct propagation path that reaches the microphone from its direction of maximum directivity. In Figure \ref{fig:t30}, we can see how the T30 also presents some small variations and, while some of these can be explained due to noisy estimates of the decay rates, the rest is explained due to the non-uniformity of the room.

\vspace{-3mm}
\section{Source Separation Experiments}
\label{sec:experiments}

In order to analyze some of the possibilities of the dataset and better understand its properties, we present a couple of experiments where we use our dataset to train X-UMX-like models for sound source separation. In both cases, we use the Tchaikovsky piece of the dataset to train the models, since it contains more different instruments, and the Mozart piece to evaluate the results. 

Since only one music piece does not contain enough diversity in terms of overlapping notes and instruments to train a separation model that generalizes to other pieces, we created \textcolor{black}{cacophonic mixtures} by taking random segments of every instrument and mixing them instead of only taking synchronized segments. This probably makes the separation problem simpler since it breaks the temporal coherence between the different instruments and their harmonic relationships, but it has been shown to improve the generalization of \textcolor{black}{MSS} models when data is scarce \cite{jeon2024does}.

As model architecture, we chose X-UMX \cite{Sawata2021} due to its simplicity and baseline vocation, having been used as baseline of the ISMIR 2021 Music Demixing (MDX) Challenge \cite{Mitsufuji2022music} and the work presenting the SynthSOD dataset \cite{Garcia2025}. This model uses the magnitude spectrogram of the input mixture to predict a \textcolor{black}{real-valued} spectral mask that is applied to the original spectrogram to generate the separated stems. The original UMX model \cite{stoter2019open} consisted of an independent branch for each stem with an encoder (composed of a linear layer with batch normalization and hyperbolic tangent activation), 3 BLSTM layers, and a decoder (composed of two linear layers with batch normalization and ReLU activation in the first one). The X-UMX architecture adds a bridging operation between the encoder and the recurrent layers and between the recurrent layers and the decoder, where the latent representations of every branch are averaged so they can share information between them. We trained the models using the combination loss presented with the X-UMX architecture and with the hyperparameters of the original Asteroid implementation \cite{Pariente2020Asteroid}.

\textcolor{black}{In addition to X-UMX, we conducted experiments using DTTnet \cite{chen_dttnet_2024}. DTTNet is a lightweight architecture that operates in the STFT domain. It integrates Dual-Path processing with a Time–Frequency Convolutional Time-Distributed Fully Connected U-Net (TFC-TDF U-Net) backbone. Notably, it achieves competitive separation performance on MUSDB18 with substantially fewer parameters than several recent approaches. For training, we used the same hyperparameters reported by the authors for the separation of the stem labeled as "other" in the MUSDB18 dataset and optimized the model using waveform L1 loss.}

Both experiments were evaluated using the fourth version of the museval library \cite{museval1}. This library computes the metrics in one-second frames and then averages them using a median operation. For the computation of the SDR, it does not allow any kind of linear distortion on the ground truths (so it is equivalent to a classical SNR \cite{le2019sdr}) but, for the computation of the signal-to-interferences ratio (SIR) and signals-to-artifacts ratio (SAR), it allows linear distortions modeled with time-invariant filters with 512 taps \cite{vincent2006performance}. We aggregated the metrics for every track by taking the median value of its frames but, since the definition of the ratios diverges when the reference signal approaches to zero, we excluded the frames where the energy of the reference signal was \SI{30}{dB} below the most energetic frame of that source in that track.

\textcolor{black}{We note that SDR-based metrics do not always correlate well with perceptual quality, as discussed in \cite{Torcoli2021}. While higher SDR values generally corresponded to audible interference reduction in our experiments, especially in the debleeding task, some perceptually relevant artifacts were not captured by these measures. Therefore the results should be interpreted as indicators of interference suppression rather than perceived audio quality. A dedicated perceptual evaluation is out of the scope of this paper.}

In this section, we present two different experiments: one in which we train a MSS model to separate the orchestra families from the stereo mix and one where we train independent models to remove the bleeding from every instrument microphone.

The code for replicating all the experiments presented in this paper is openly available in our official repository\footnote{https://github.com/repertorium/TheSpheresDataset-Experiments}.

\subsection{Family separation}
\label{sec:separation}

\begin{table*}
    \caption{Signal-to-distortion ratio (SDR), signal-to-interferences ratio (SIR), signals-to-artifacts ratio (SAR), and Source Image-to-Spatial distortion Ratio (ISR) for every orchestra family in the original Mozart piece of The Spheres dataset and the Operation Beethoven track and the results obtained by a X-UMX separation model trained on the Tchaikovsky piece of The Spheres dataset, \textcolor{black}{SynthSOD, and SynthSOD plus finetuning in the Tchaikovsky piece of The Spheres dataset}.}
    \centering
    \sisetup{table-format=2.1, table-auto-round}
    \begin{tabular}{llSSSSSSS}
    \toprule
    \multirow{4}{*}{Eval dataset}                   & \multirow{4}{*}{Source}   & {\multirow{2}{*}{Original}}   & \multicolumn{6}{c}{Training dataset}                                                  \\ 
                                                                                                                \cmidrule(lr){4-9} 
                                                    &                           &                               & \multicolumn{4}{c}{Tchaikovsky (The Spheres)} & {\hspace{6pt}SynthSOD\hspace{6pt}} & {Synth.+Tchaik.}  \\    
                                                                                \cmidrule(lr){3-3}              \cmidrule(lr){4-7}                              \cmidrule(lr){8-8}  \cmidrule(lr){9-9}
                                                    &                           & {SDR }                        & {SDR}     & {SIR}     & {SAR}     & {ISR}     & {SDR}             & {SDR}             \\
                                                    &                           & {[dB]}                        & {[dB]}    & {[dB]}    & {[dB]}    & {[dB]}    & {[dB]}            & {[dB]}            \\
    \midrule
    \multirowcell{3}[0pt][l]{Mozart\\ (The Spheres)}& Strings                   &  4.48                         &  9.44     &  11.93    & 12.08     & 18.24     & 6.84              & 9.11              \\
                                                    & Woodwinds                 & -6.32                         &  3.72     &   9.91    &  3.44     &  7.48     & 1.79              & 3.64              \\
                                                    & Brass                     & -9.18                         &  0.78     &   4.43    &  1.01     &  7.19     & 0.03              & 0.76              \\
    \midrule
    \multirowcell{4}[0pt][l]{Operation\\ Beethoven} & Strings                   &  2.09                         &  2.78     &   5.57    &  1.42     &  3.77     & 1.34              & 0.89              \\
                                                    & Woodwinds                 & -7.70                         &  0.25     &  -6.40    &  0.12     &  2.10     & 1.06              & 0.04              \\
                                                    & Brass                     & -9.54                         & -2.92     & -15.83    &  0.43     &  6.69     & 0.13              &-3.94              \\
                                                    & Percussion                & -2.74                         &  2.58     &  20.64    &  0.16     & -5.35     & 0.02              & 2.66              \\
    \bottomrule
    \end{tabular}
    \label{tab:sep_results}
\end{table*}

Due to the high difficulty of separating every instrument of the orchestra from a stereo mix, we designed an easier experiment where the instruments were grouped according to their orchestra family (strings, woodwinds, brass, and percussion) and trained an X-UMX model with 4 branches to separate them.

Table \ref{tab:sep_results} shows the original SDR, SIR, and SAR, in the stereo mix of the Mozart piece of The Spheres and the recording from the Operation Beethoven dataset and the results obtained with the X-UMX model trained on the Tchaikovsky piece of The Spheres. The SIR of the separated stems shows important improvements in terms of source separation for all the families in the Mozart piece, although the weaker the family was in the original mixture, the higher are the artifacts generated by the model (it should be noted that the Mozart piece does not contain any percussion instruments and only the horn from the brass family). However, we can see that the model does not generalize to other datasets such as Operation Beethoven, where Woodwinds and Brass show a clear degradation in SIR, and all instrument families exhibit a high level of artifacts. Several differences exist between both datasets, such as the room, the instrument and microphone positions, and the instruments and microphones used; further research would be needed to analyze the importance of these factors in the generalization of \textcolor{black}{MSS} models. \textcolor{black}{Nevertheless, we can conclude that generalizing to different recording environments is a much more challenging task than generalizing to different pieces recorded under the same conditions.}

\textcolor{black}{Additionally, we also trained a X-UMX model for family separation on the SynthSOD dataset, which we resampled at \SI{48}{kHz} to match the sampling rate of The Spheres, and tried to finetune it using the Tchaikovsky piece of The Spheres. In Table \ref{tab:sep_results} we can see how the model trained using synthetic material does not generalize well to real recordings. In the case of the finetuned model, it does not seem to obtain any benefits from the pretraining on SynthSOD, and it obtains equivalent results to the model trained solely on the Tchaikovsky piece. Despite pretraining and finetuning strategies can potentially be useful for MSS \cite{Sarkar2023}, more sophisticated approaches need to be developed for cases of higher polyphony and acoustic complexity.}

\subsection{Microphone debleeding}
\label{sec:debleeding}

\begin{table}
    \centering
    \footnotesize
    \setlength{\tabcolsep}{3pt}
    \sisetup{table-format=2.1, table-auto-round}
    \caption{Signal-to-distortion ratio (SDR), signal-to-interferences ratio (SIR), signal-to-artifacts ratio (SAR), and Source Image-to-Spatial distortion Ratio (ISR) of the close microphones of the Mozart piece (The Spheres) and the results obtained by \textcolor{black}{X-UMX and DTTnet} debleeding models trained on the Tchaikovsky piece.}
    \begin{tabular}{lSSSSSSSSS}
\toprule
\multirow{3}{*}{Source} & {Original} & \multicolumn{4}{c}{Separated X-UMX} & \multicolumn{4}{c}{Separated DTTnet} \\
\cmidrule(lr){2-2}\cmidrule(lr){3-6}\cmidrule(lr){7-10}
& {SDR} & {SDR} & {SIR} & {SAR} & {ISR} & {SDR} & {SIR} & {SAR} & {ISR} \\
& {[dB]}& {[dB]}& {[dB]}& {[dB]}& {[dB]}& {[dB]}& {[dB]}& {[dB]}& {[dB]}\\
\midrule
Violin I      & -2.44 & 2.40  & 10.68 & 1.38  & -0.47 & 3.547 & 5.263  & 5.575  & 14.901 \\
Violin I+II   & 1.40  & 7.61  & 14.18 & 8.20  & 15.04 & 7.900 & 16.322 & 8.222  & 12.991 \\
Violin II     & -7.01 & 0.96  & 7.99  & -1.01 & 2.81  & 0.209 & 7.811  & 0.012  & 2.598 \\
Violin II+I   & -4.27 & 4.14  & 11.13 & 3.39  & 8.18  & 4.484 & 9.679  & 4.674  & 10.674 \\
Viola         & -2.81 & 4.31  & 11.72 & 3.63  & 7.88  & 3.483 & 11.867 & 2.945  & 9.024 \\
Cello         & -2.45 & 4.26  & 10.69 & 4.17  & 9.14  & 3.451 & 10.184 & 3.338  & 8.519 \\
Bass I+II     & 5.70  & 10.06 & 18.83 & 10.96 & 14.95 & 11.251& 17.813 & 13.165 & 17.152 \\
Bass II+I     & 11.34 & 16.09 & 23.18 & 17.29 & 22.42 & 15.611& 23.768 & 17.982 & 20.749 \\
Flute         & -1.75 & 11.54 & 20.54 & 11.28 & 19.77 & 11.241& 21.941 & 11.581 & 16.117 \\
Clarinet      & 2.76  & 6.80  & 17.43 & 6.33  & 9.55  & 5.970 & 19.296 & 5.067  & 8.926 \\
Oboe          & -3.76 & 4.74  & 9.89  & 5.41  & 12.46 & 4.973 & 11.199 & 4.893  & 11.483 \\
Bassoon       & -0.19 & 7.08  & 16.12 & 6.64  & 10.62 & 7.997 &  17.510  & 7.437 & 12.856 \\
Horn          & 11.86 & 14.60 & 21.50 & 16.13 & 20.29 & 13.701& 24.563 & 13.878 & 15.015 \\
\bottomrule
\end{tabular}
\label{tab:deb_results}
\end{table}

An easier but quite interesting application of \textcolor{black}{MSS} for classical music production is debleeding the close microphone signals. Orchestras are typically recorded using microphones close to every instrument section apart from the main stereo microphone, and those signals are used in the final stereo mix to increase the presence of some instruments that might be too weak in the stereo microphone. However, the signals of these close microphones contain a high level of bleeding from all the other instruments of the orchestra, making the work of the mixing engineer nontrivial. Therefore, having models capable of removing this bleeding and ensuring that the signal of the close microphones contain only the sound of its corresponding instrument would be of great interest to the music-recording industry. \textcolor{black}{Unlike previous resources, The Spheres dataset provides close-microphone recordings for each instrument alongside the simultaneous leakage, or bleeding, present in every other microphone's signal.}

We trained individual one-branch X-UMX \textcolor{black}{and DTTnet} models for every instrument/section microphone using the Tchaikovsky piece of The Spheres. As input, we took the mix of the signals from all the instruments at the corresponding microphone and, as target output, we took the signal of the instrument that was intended to be recorded by that microphone, so we can see the debleeding problem as an enhancement task. More advanced architectures could be used to include the signals from the other microphones as inputs of the model so it has a reference of the bleeding that it needs to remove, but we decided to keep the model as simple as possible since this is a preliminary experiment and the goal is to showcase some of the applications and properties of the dataset.
 
\textcolor{black}{As shown in Table \ref{tab:deb_results}, all models successfully improved the SDR of the target instruments. Notably, the models achieved substantial interference reduction, with SIR gains ranging from $5$~dB to $24.6$~dB, while maintaining reasonable SAR levels (generally above $3$~dB, with the exception of the violins).}
Debleeding the microphones of the violin I and violin II sections is especially difficult, since they contain the bleeding from the other violin section that has the same timbral properties. To facilitate this case, we also trained a model to remove all the bleeding from the violin I microphone except the one coming from the violin II section (Violin I+II in Table \ref{tab:deb_results}) and vice versa (Violin II+I); note that this is the only approach possible for the Bass I and II microphones, since we do not have the isolated recordings of every bass and they played the same melody.

Since The Spheres is the first public dataset to include close microphone signals, we could not evaluate our debleeding models trained on the Tchaikovsky piece against an external dataset, as was possible for the family-separation experiments with the Operation Beethoven recording. Nevertheless, we had access to raw material from private professional sessions by the studio’s owner (Odradek Records) which, although lacking isolated tracks for objective evaluation, allowed us to subjectively confirm the same generalization issues discussed in Section~\ref{sec:separation}. \textcolor{black}{As for the family separation, models trained with the Tchaikovsky piece of The Spheres dataset seem to be able to obtain good results in other music pieces recorded under the same conditions, but fail to generalize to different recording environments.}

\subsection{Discussion}

One important conclusion can be drawn from the previous experiments: training separation models with signals from a single recording setup does not guarantee generalization to other setups and, therefore, models should not be trained and tested in the same setup if there is not a justified use case that makes feasible the recording of the training data in the same conditions where the models will be used. In the previous experiments, we trained our models in the Tchaikovsky piece of The Spheres and then evaluated them in the Mozart one since our main goal was the study of the properties of the dataset, but we discourage researchers from doing this in further research works.

We believe that the main interest of this dataset is in the evaluation of the models trained with different datasets, not in using it for training. The Spheres is the first dataset that includes microphones close to every instrument section following the same arrangement that is used in real orchestra recordings, opening the door to the evaluation of debleeding models, which is a problem that has barely been studied but has a big importance for the music recording industry.

\textcolor{black}{From the presented results, we can conclude that the generalization to unseen recording environments is more challenging than the generalization to unseen repertoire. The models trained on the Tchaikovsky piece were able to obtain moderate or good results in the Mozart piece despite the low amount of training material (which was augmented using cacophonic remixes), but failed when applied on completely different recordings. This suggests that, for the development of future datasets, if they want to be useful for training, they should focus more on the diversity of the recording conditions than in the absolute duration of the dataset or in the number of music pieces.}

An interesting approach that has not been studied in this paper is the use of the signals from the scales and solos included in The Spheres to finetune models trained with larger synthetic datasets such as SynthSOD \cite{Garcia2025}. Having the references of every instrument for a full music piece (as the Tchaikovsky piece included in The Spheres) with the same recording setup in which the model is going to be used is not a realistic assumption for most real-world applications, but recording some small extracts from every instrument section (as the scales and solos included in The Spheres) might be possible during the rehearsals of the orchestra where the separation system is going to be used. \textcolor{black}{However, more sophisticated pretraining and finetuning techniques should be developed for this.}

\vspace{-3mm}
\section{Conclusion}

In this paper, we present The Spheres dataset, the first publicly available multitrack orchestral recordings capturing classical music through a comprehensive set of microphone signals, including both main and close spot microphones. Unlike previous datasets, this collection provides recordings as isolated stems for each instrument section from all microphone positions, accurately reflecting the conditions of real orchestral productions and offering isolated signals for research purposes.

The dataset openly publishes two complete canonical works—Tchaikovsky’s Romeo and Juliet and Mozart’s Symphony No. 40—as well as solo scales with different dynamics and playing techniques. In addition, Room Impulse Responses (RIRs) were measured for every instrument position and are released alongside the dataset, providing a detailed characterization of the recording space. We have also provided analyses of the data, including polyphony levels, instrument activity times, and acoustic properties derived from the RIRs, which illustrate both the diversity and complexity of the material. Furthermore, we performed experiments on instrument-family separation from the main stereo microphones and on microphone debleeding at the spot level. While the \textcolor{black}{analyzed} models improved separation in several cases, they also confirmed the difficulty of orchestral source separation \textcolor{black}{and showcase the need for further research on MSS methods that can generalize to different recording conditions}.

Overall, The Spheres dataset represents a significant step forward for Music Source Separation (MSS) in the classical domain. By offering real, professionally recorded multitrack material with aligned main, ambient, and spot microphones, it provides the community with a unique benchmark for evaluating separation, dereverberation, localization, and immersive rendering approaches. We expect this dataset to become a valuable resource for future research, helping to bridge the gap between synthetic data training and real-world orchestral applications.

\vspace{-3mm}
\appendices
\section{}\label{appendix:ESS}

This appendix provides the mathematical background for estimating RIRs using the exponential sine sweep (ESS) technique and details a modified procedure developed to address corruption in the reference sweep signal.

Estimating the RIR with ESS involves exciting the system with a test signal defined as:

\begin{equation}\label{eq:ess}
    x(t) = sin\left(\frac{2\pi f_1 T}{ln\left(\frac{f_2}{f_1}\right)}\left[e^{\frac{t}{T}ln\left(\frac{f_2}{f_1}\right)}-1\right]\right)
\end{equation}

where $f_1$ and $f_2$ are the starting and stopping sweep frequencies measured in Hz, $T$ is the sweep duration in seconds and $ln\left(\frac{f_2}{f_1}\right)$ is known as the sweep rate. These parameters define the ESS design.

An inverse filter $x_{inv}(t)$ is designed so that its convolution with $x(t)$ approximates a delayed Dirac delta:

\begin{equation}\label{eq:ess_inv_property}
    x(t) * x_{inv}(t) \approx \delta(t - t_0)
\end{equation}

where $*$ denotes linear convolution, $\delta(t)$ is the Dirac delta function, and $t_0$ is a time delay. The closed-form expression of the inverse filter is:

\begin{equation}\label{eq:inv_filter}
    x_{inv}(t) = x(T-t)e^{-\frac{t}{T}ln\left(\frac{f_2}{f_1}\right)}
\end{equation}

The inverse filter described in Eq. $\ref{eq:inv_filter}$ is computed by time reversing the original ESS, $x(t)$, delaying the result to obtain a casual signal and apply an amplitude modulation with a 6 dB/octave gain \cite{Delvaux2014USINGAE}.

By convolving each recorded microphone signal $y(t)$ with this inverse filter $x_{inv}(t)$, the RIR can be extracted:

\begin{equation}\label{eq:rir_estimation}
    r(t) = y(t) * x_{inv}(t) = h(t-t_0) + \eta(t)
\end{equation}

where $h(t)$ is the delayed linear RIR, and $\eta(t)$ contains higher-order distortion products that appear earlier in time and can be separated via time-windowing.

In our measurements, the reference ESS signal was corrupted from improper time-stretching and downsampling, which introduced spectral distortion and phase misalignment relative to the designed ESS parameters. Consequently, the original analytical inverse could not be used and a data-driven inverse filter was required. To address this, we computed an equivalent inverse filter in the frequency domain. Let $x[n]$ be the discrete, corrupted ESS signal and $\delta[n-n_0]$ a delayed discrete delta function. The inverse filter was computed as:

\vspace{-2mm}
\begin{equation}\label{eq:freq_domain_inv}
x_{inv}[n] = \Re\left\{\mathcal{DFT}^{-1}\left\{\frac{\mathcal{DFT}\{\delta[n-n_0]\}}{\mathcal{DFT}\{x[n]\}}\right\}\right\}
\end{equation}
where $\Re$ indicates real part, $\mathcal{DFT}$ and $\mathcal{DFT}^{-1}$ denote the DFT and IDFT operators respectively.

\begin{figure}
    \centering
    \includegraphics[width=\columnwidth]{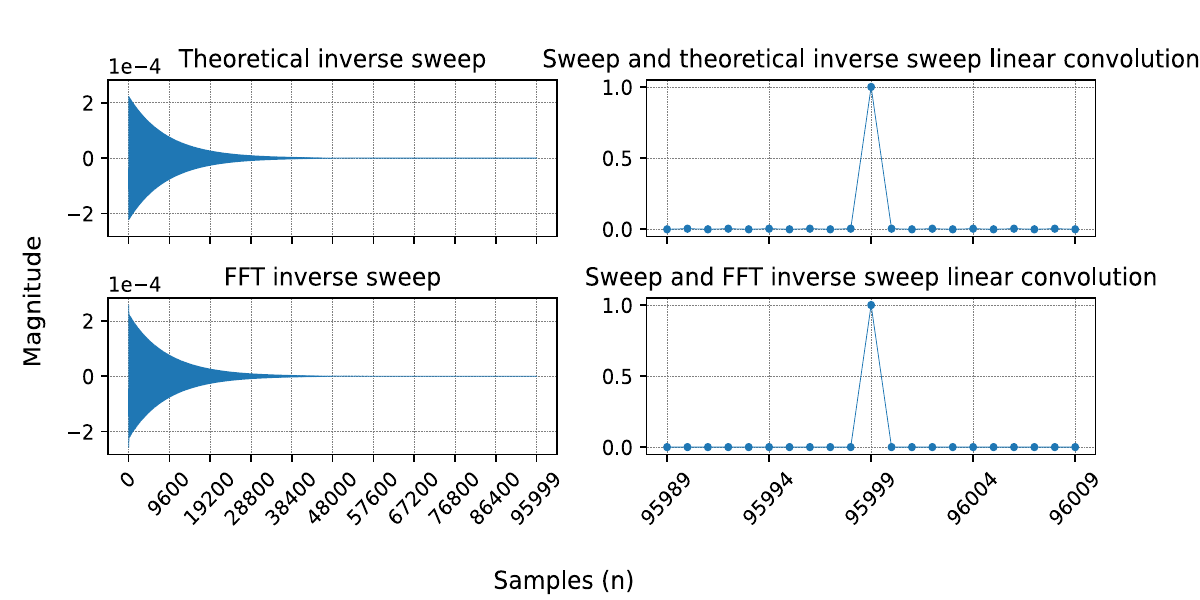}
    \caption{Comparison between the theoretical and proposed inverse filters. Top-left: Theoretical inverse filter derived from known ESS parameters. Top-right: Linear convolution of the reference sweep with the theoretical inverse filter, showing the expected delayed impulse. Bottom-left: Inverse filter computed in the frequency domain as described in eq. \ref{eq:freq_domain_inv}. Bottom-right: Convolution result using the proposed inverse filter, demonstrating preservation of the delayed impulse property in eq. \ref{eq:ess_inv_property}.}
    \label{fig:inverse_ess_time}
\end{figure}

To validate that the inverse filter computed in the frequency domain preserves the key convolution property in Eq. \ref{eq:ess_inv_property}, we compared it against the theoretical inverse filter derived from known ESS parameters. The reference sweep design parameters (see \ref{eq:ess}) are $f_1 = 1$ Hz, $f_2 = 48000$ Hz, $T = 1$~s, with a sampling rate of $96000$ Hz. Figure~\ref{fig:inverse_ess_time} shows both inverse filters and their corresponding convolution results with the reference sweep.

\begin{figure}
    \centering
    \includegraphics[width=\columnwidth]{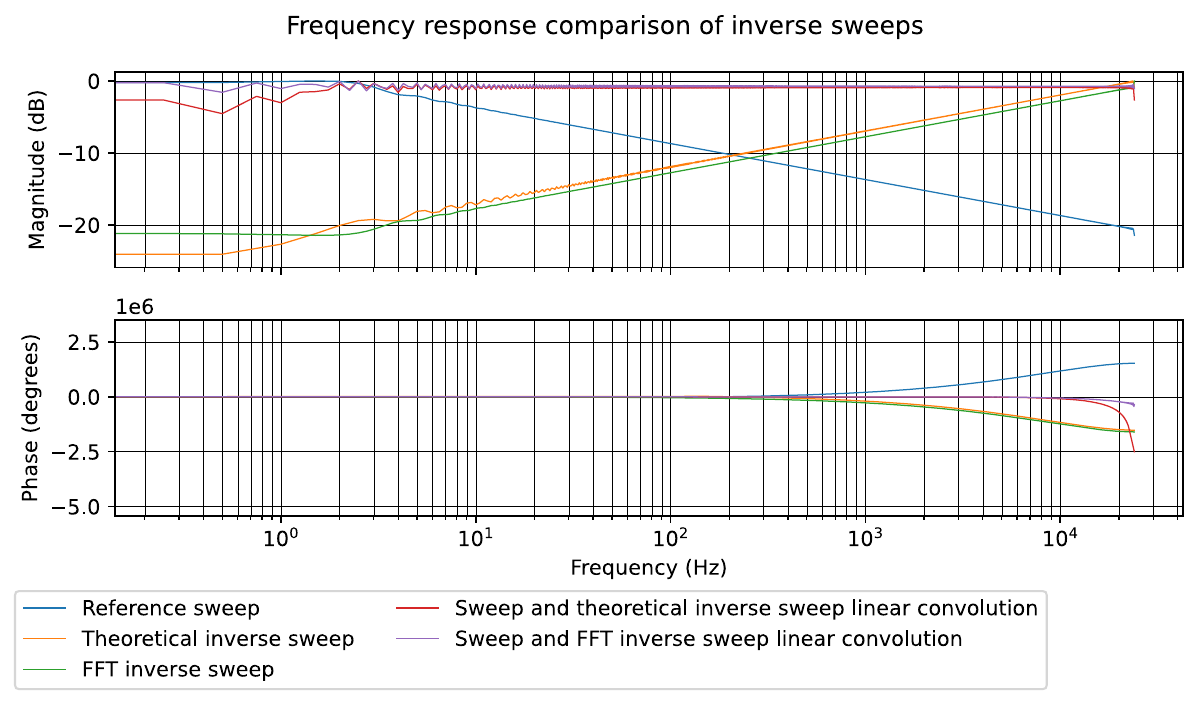}
    \caption{Frequency-domain validation of the proposed inverse filter. The plots show the Bode-style magnitude (top) and phase (bottom) responses of: the reference ESS (blue), the theoretical inverse filter (orange), the proposed frequency-domain inverse filter (green), and the convolution results for both inverse filters (red and purple). The magnitude response is flat across the entire frequency range, confirming accurate compensation of the sweep spectrum. Minor phase differences are observed but are consistent between the two, demonstrating that the proposed inverse filter preserves the desired convolution property.}
    \label{fig:inverse_ess_freq}
\end{figure}

\begin{figure}
    \centering
    \includegraphics[width=\columnwidth]{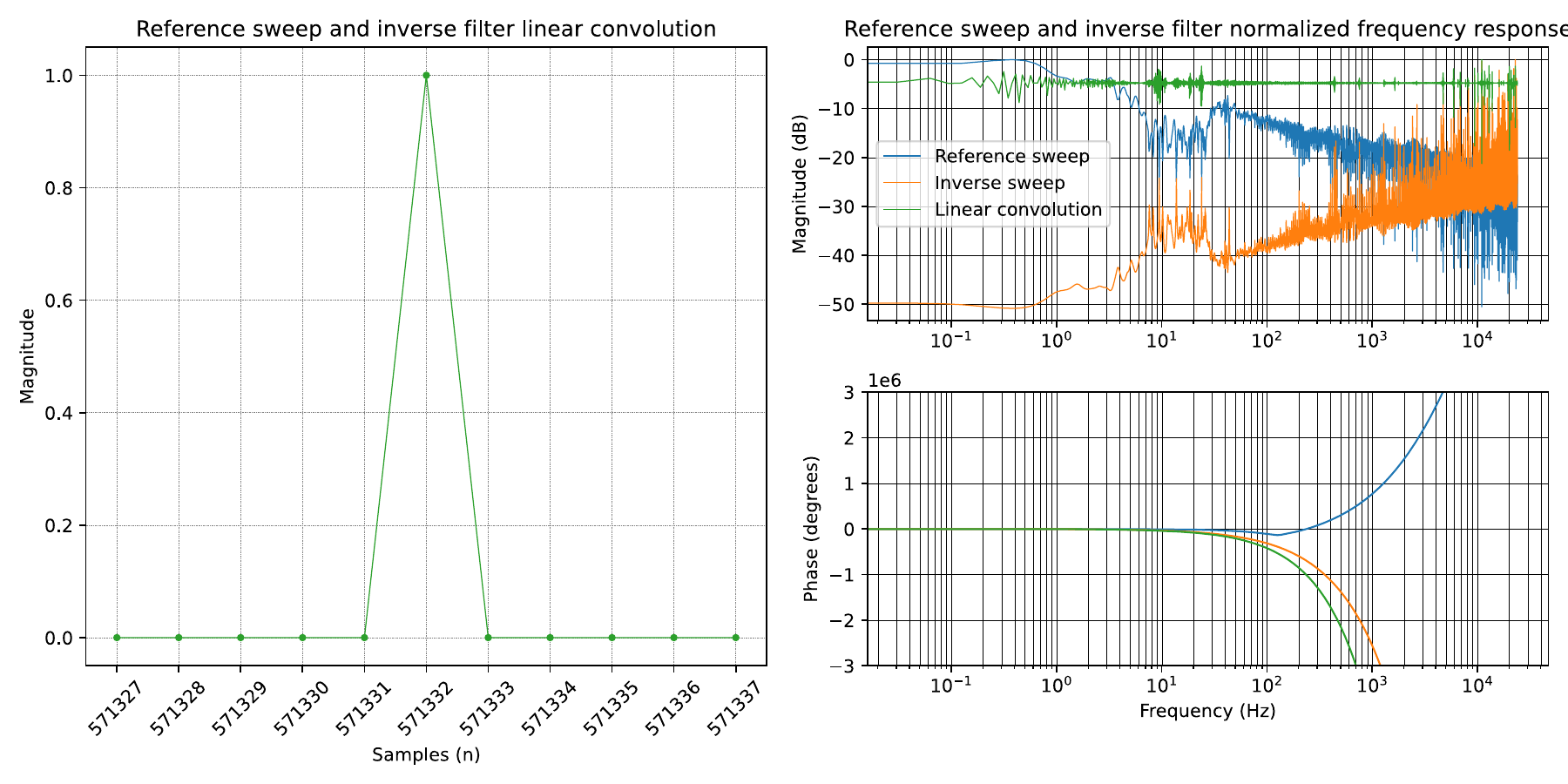}
    \caption{Inverse filter computation and validation for the corrupted ESS used in the actual measurements. Left: Result of the linear convolution between the corrupted reference sweep and its computed inverse filter, showing the expected delayed impulse. Top-right: Magnitude responses of the corrupted sweep (blue), the computed inverse filter (orange), and their convolution result (green). The green trace is generally flat but exhibits some frequency-dependent ripple, which may introduce slight coloration into the estimated RIR. Bottom-right: Corresponding phase responses.}
    \label{fig:computed_ess}
\end{figure}

Figure~\ref{fig:computed_ess} shows the computed inverse filter for the corrupted sweep used in the actual room measurements. Since the sweep’s design parameters are unknown, the theoretical inverse filter cannot be obtained for direct comparison. Nevertheless, the convolution result (left) exhibits a sharp, delayed impulse, and the magnitude response of the convolution (green, top-right) is generally flat, confirming that the proposed approach remains effective. Minor ripple in the magnitude response indicates some expected coloration in the estimated RIR. The phase response (bottom-right) closely resembles those presented in Figure~\ref{fig:inverse_ess_freq}, reinforcing the validity of the frequency-domain inverse filter for practical use.

\section*{Acknowledgment}

This work was supported by “REPERTORIUM” Project under Grant Agreement 101095065. Horizon Europe. Cluster II. Culture, Creativity and  Inclusive Society. Call HORIZON-CL2-2022-HERITAGE-01-02. \\ 
The authors wish to acknowledge CSC—IT Center for Science, Finland, for computational resources. \\
The authors thank the Colibrì Ensemble musicians for their participation and
for granting consent for the use and open distribution of their recordings
for research purposes.\\

\ifCLASSOPTIONcaptionsoff
  \newpage
\fi

\vspace{-3mm}
\bibliographystyle{IEEEtran}

\bibliography{IEEEabrv,References,References_extra}

\vfill\pagebreak

\end{document}